%% file: arxiv.tex
\documentclass[10pt,conference,letterpaper]{IEEEtran}
\IEEEoverridecommandlockouts
\pdfoutput=1
\usepackage[T1]{fontenc}
\usepackage{cite}
\usepackage{amsmath,amssymb,amsfonts}
\usepackage{algorithm}
\usepackage[noend]{algpseudocode}
\usepackage{flushend}
\usepackage{graphicx}
\usepackage{textcomp}
\usepackage{dcolumn} 
\usepackage{bm} 
\usepackage{braket}
\usepackage{xspace}
\usepackage{hyperref} 
\hypersetup{
    colorlinks = true,
    citecolor = magenta,
    linkcolor = purple
}
\def\BibTeX{{\rm B\kern-.05em{\sc i\kern-.025em b}\kern-.08em
    T\kern-.1667em\lower.7ex\hbox{E}\kern-.125emX}}

\usepackage{comment}

\input{custom-command}
    
\begin{document}
\bstctlcite{IEEEexample:BSTcontrol}

\title{Cross-Validating Quantum Network Simulators
\thanks{J.C. and M.H. contributed equally. This work was supported by JST [Moonshot R\&D Program] Grant
Number [JPMJMS226C].
This material is based upon work supported by the U.S. Department of Energy Office of Science National Quantum Information Science Research Centers as part of the Q-NEXT center.
}
}

\author{
\IEEEauthorblockN{
Joaquin Chung\IEEEauthorrefmark{1}\IEEEauthorrefmark{2},
Michal Hajdu\v{s}ek\IEEEauthorrefmark{3},
Naphan Benchasattabuse\IEEEauthorrefmark{3},
Alexander Kolar\IEEEauthorrefmark{1}\IEEEauthorrefmark{2},
Ansh Singal\IEEEauthorrefmark{1}\IEEEauthorrefmark{5},\\
Kento Samuel Soon\IEEEauthorrefmark{3},
Kentaro Teramoto\IEEEauthorrefmark{3}\IEEEauthorrefmark{4},
Allen Zang\IEEEauthorrefmark{1}\IEEEauthorrefmark{2},
Raj Kettimuthu\IEEEauthorrefmark{1}\IEEEauthorrefmark{2}, and
Rodney Van Meter.\IEEEauthorrefmark{3}}

\IEEEauthorblockA{\IEEEauthorrefmark{1}\textit{Argonne National Laboratory, USA},~\IEEEauthorrefmark{2}\textit{The University of Chicago, USA},~\IEEEauthorrefmark{3}\textit{Keio University, Japan}}
\IEEEauthorblockA{\IEEEauthorrefmark{4}\textit{Mercari Inc., Japan},~\IEEEauthorrefmark{5}\textit{Northwestern University, USA}\\
emails: \{chungmiranda, akolar, asingal, yzang, kettimut\}@anl.gov, \{michal, whit3z, soon, zigen, rdv\}@sfc.wide.ad.jp
}
}

\thispagestyle{plain}
\pagestyle{plain}
\maketitle

\begin{abstract}
We present a first cross-validation of two open-source quantum network simulators, QuISP and SeQUeNCe, focusing on basic networking tasks to ensure consistency and accuracy in simulation outputs. 
Despite very similar design objectives of both simulators, their differing underlying assumptions can lead to variations in simulation results. 
We highlight the discrepancies in how the two simulators handle connections, internal network node processing time, and classical communication, resulting in significant differences in the time required to perform basic network tasks such as elementary link generation and entanglement swapping.
We devise common ground scenarios to compare both the time to complete resource distribution and the fidelity of the distributed resources.
Our findings indicate that while the simulators differ in the time required to complete network tasks—a constant factor difference attributable to their respective connection models—they agree on the fidelity of the distributed resources under identical error parameters.
This work demonstrates a crucial first step towards enhancing the reliability and reproducibility of quantum network simulations, as well as leading to full protocol development.
\rdv{not just sims but full protocol development}
Furthermore, our benchmarking methodology establishes a foundational set of tasks for the cross-validation of simulators to study future quantum networks.
\end{abstract}

\maketitle

\section{Introduction}
\label{Introduction}

Quantum networks~\cite{van2014quantum} hold great promise for enhancing existing functionality of classical communication networks~\cite{ekert1991quantum,gottesman2012longer,komar2014quantum}, as well as enabling new applications~\cite{wang:RFC9583,broadbent2009universal,cuomo2020towards}, and they are widely expected to be necessary for scalable, fault-tolerant quantum computers~\cite{simmons2024scalable,PRXQuantum.2.017002,koyama2024optimal}.
However, the most ambitious goal is to create a quantum Internet~\cite{wehner2018quantum} connecting quantum resources across the globe.

\begin{figure}[t]
    \centering
    \includegraphics[width=\linewidth]{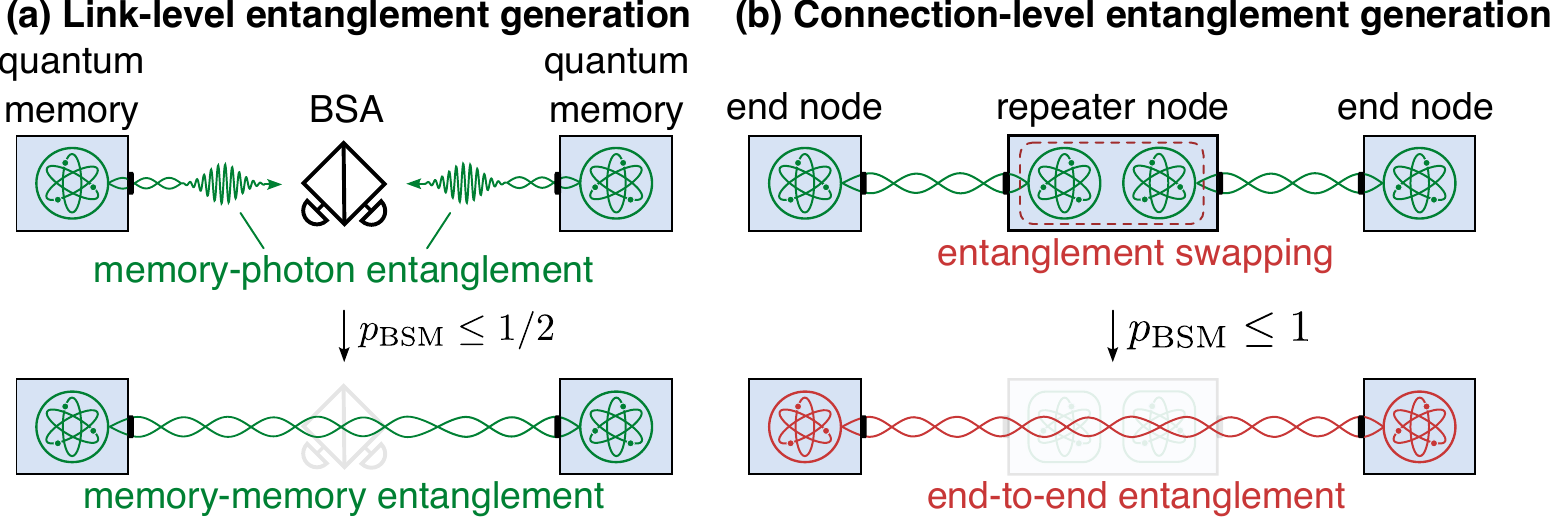}
    \caption{(a) Entanglement generation using photonic \textit{Bell-state measurement} (BSM) at the \textit{Bell-state analyzer} (BSA) in the \textit{memory-interference-memory} (MIM) link architecture. Success probability is denoted by $p_{\text{BSM}}$. (b) After successful generation of two neighboring links, the repeater performs entanglement swapping on its quantum memories to create an end-to-end entangled connection.}
    \vspace{-0.2in}
\label{fig:preliminaries}
\end{figure}

Quantum networks are fundamentally different from their classical counterparts.
First, quantum-networked applications consume entangled states as a resource~\cite{hajdusek2023quantum}.
Second, the attenuation in optical fiber is a major source of loss since quantum signals cannot be reliably amplified or copied due to the no-cloning theorem~\cite{Park1970,dieks1982communication,zurek1982single}. 
To overcome these design constraints, quantum repeaters have been introduced~\cite{briegel1998quantum,azuma2023quantum}.
Generally, the link-level entanglement generation is non-deterministic~\cite{lutkenhaus1999bell} and requires classical messages to herald success or failure.
Upon successful generation of link-level entanglement between adjacent nodes of a quantum repeater, entanglement swapping (ES)~\cite{PhysRevLett.71.4287,jin2015highly,hermans2022qubit} splices two entanglement links into a long-range entangled state shared between the desired end nodes of the quantum network.
This highlights the fact that remote entanglement generation is akin to coordinated and distributed computation~\cite{vanmeter2022quantum,kozlowski2023rfc}, where quantum repeaters send and receive a constant stream of quantum and classical messages and execute actions based on both local and non-local conditions.
The mechanism of a quantum repeater is visualized in Fig.~\ref{fig:preliminaries}.

The study of quantum networks quickly becomes intractable using analytical methods as we scale up the system.
Therefore, we must rely on simulations.
Ideally, analysis, simulation, and experiment agree over a broad enough range of conditions to give us confidence in the predictions of the simulator.
Recent years have seen the introduction of a number of quantum network simulators, including NetSquid~\cite{coopmans2021netsquid}, QuISP~\cite{satoh2022quisp}, QuNetSim~\cite{diadamo2012qunetsim}, QuNet~\cite{leone2021qunet}, ReQuSim~\cite{wallnofer2024faithfully}, and SeQUeNCe~\cite{sequence}.
Despite overlapping aims and goals, these simulators differ significantly.
However, the results of the simulations should agree.
When the simulators produce different results on similar configurations, we must consider three possible cases: (a) real differences exist due to design choices in the network, (b) valid but differing simplifications in simulation were made at the physical or protocol level, or (c) outright bugs in one or possibly both simulators exist.
Case (a) represents the differences we wish to study as we commit our design decisions on the way to real-world implementations.
Case (b) represents an opportunity to refine our understanding of how to most accurately simulate quantum networks for parameters of interest. 
Case (c) should simply be fixed. 

The cases exposed above emphasize the need for cross-validation studies of existing quantum network simulators.
In this paper, we present a comparative study of two quantum network simulators: QuISP and SeQUeNCe.
By focusing on simulation of fundamental quantum networking primitives such as link-level entanglement generation and entanglement swapping, we identify the discrepancies in how the simulated protocols handle connections, internal node processing time, and classical communication overhead.
We compare the simulators in terms of the predicted time needed to generate a set number of entangled states between two end nodes as well as the fidelity of these distributed states.
We find that the simulators disagree on the predicted time, differing by a constant factor.
Using the differences in their designs, we explain and correctly quantify this offset.
Furthermore, the simulators agree on the predicted fidelities of the final states, pointing to correct implementation of realistic error models in both simulators.
Finally, the tests and benchmarking methodologies presented here represent an important step in developing a full set of cross-validation techniques applicable across all simulators of quantum networks.

We first describe the two simulators in Sec.~\ref{Simulators}.
Then we detail the three experiments executed for simulator cross-validation in Section~\ref{Benchmarking}.
Before running the experiments, we develop detailed models for the timing of classical and quantum messages in the simulators as well as discuss their error models in Sec.~\ref{Models}.
We present the results in Sec.~\ref{Results}, and conclude with discussion in Sec.~\ref{Discussion}.

\section{The Simulators}
\label{Simulators}

\begin{figure}
    \centering
    \includegraphics[width=\linewidth]{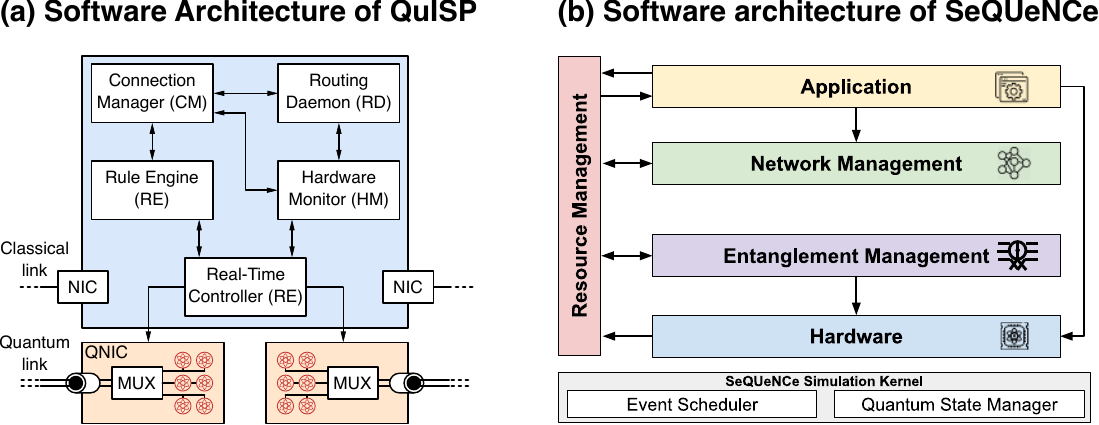}
    \caption{(a) Quantum Router Software Architecture (QRSA) used in QuISP to control the behavior of network nodes. (b) SeQUeNCe software framework comprises a simulation kernel and five other modules.}
    \vspace{-0.2in}
\label{fig:software_arch}
\end{figure}

\emph{QuISP---} Quantum Internet Simulation Package (QuISP)~\cite{satoh2022quisp} is an open-source simulator for large-scale quantum networks and internetworks.
It is built on top of OMNeT++~\cite{varga2010omnetpp}, a discrete-event C++ library for classical network simulators.
The primary goals of QuISP are to aid in protocol design, architecture and performance prediction, and evaluation of dynamic and emergent behavior of large scale quantum networks and internetworks using minimal computational resources.
QuISP assumes that each node of a quantum network is equipped with one or more \textit{quantum network interface cards} (QNICs) containing a number of quantum memories, which are used for storage and processing of quantum information.
Inter-node quantum communication uses single photons to establish entangled pairs between nearest-neighbor quantum network nodes.
Currently, QuISP supports three link-level architectures~\cite{jones2016design,soon2024implementation,fittipaldi2024entanglement}.
In order to control and coordinate the operation of the quantum nodes, QuISP models the \textit{Quantum Router Software Architecture} (QRSA)~\cite{satoh2022quisp} shown in Fig.~\ref{fig:software_arch}(a).
At the heart of this software architecture is the notion of a \textit{RuleSet}~\cite{matsuo2019quantum}, which consists of a list of \textit{Condition} and \textit{Action Clauses}.
The actions are triggered upon satisfaction of local conditions.
The structure of RuleSets is not fixed but can be extended and defined by the user in order to achieve the desired node behavior~\cite{satoh2023rula}.
RuleSets are generated by the \textit{Connection Manager} and executed by the \textit{Rule Engine}.
Information about physical qubits inside the QNICs is gathered by the \textit{Hardware Monitor} and passed to the Connection Manager and the \textit{Routing Daemon}.

QuISP supports a number of error models for the stationary and flying qubits.
Unitary gates and qubit measurements are also subject to errors.
Photon emission from a quantum memory and coupling to an optical fiber is assumed to be probabilistic.
Detectors used to measure flying qubits are characterized by their efficiency and dark count rates.
All of these error parameters can be defined by the user.

\emph{SeQUeNCe---} Simulator of QUantum Network Communication (SeQUeNCe)~\cite{sequence} is an open-source discrete-event simulator for detailed, accurate, and highly customizable simulations of quantum networks.
SeQUeNCe is developed in Python 3.10+ and it is designed to track billions of events per second of simulation time on a single machine.
Furthermore, a parallel version of SeQUeNCe~\cite{wu2024parallel} can achieve a speedup of up to 25$\times$ with respect to the sequential version. 
SeQUeNCe can be used to understand the trade-offs of alternative quantum network architectures, optimize quantum hardware, and study quantum networking protocols at scale. 

SeQUeNCe adopts a modularized design, shown in Fig.~\ref{fig:software_arch}(b), which enables researchers to customize the simulated network and reuse existing models in a flexible way.
The simulator is composed of five modules and a simulation kernel.
The simulation kernel provides an interface for discrete event simulation via a scheduler; it additionally tracks and updates the quantum state of simulation components via a quantum state manager.
The hardware module implements various quantum network components describing their behavior and modeling errors.
The entanglement management module includes protocols for reliable, high-fidelity distribution of entangled qubit pairs between end nodes in a quantum network. 
The resource and network management modules use classical control messages to manage allocation of local (per node) and network resources, respectively. 
Similar to QuISP, the resource manager uses an internal set of rules to manage the state of quantum memories.
The application module represents quantum network applications and their service requests. 

One goal of SeQUeNCe is to provide realism in representing quantum states to accurately track quantum states in the network.
In SeQUeNCe, quantum states can be represented using the bra-ket notation or density matrices~\cite{zang2022simulation}.
The quantum state manager provides external interfaces for quantum state manipulation. SeQUeNCe also incorporates analytical error models for efficient simulation of imperfect bipartite-entanglement distribution~\cite{zang2023entanglement,zang2024quantum}.

\section{Benchmarking Methodology}
\label{Benchmarking}
    
\begin{table}[]
\centering
\caption{Summary of the experimental settings}
\label{tab:experiments}
\resizebox{\linewidth}{!}{%
\begin{tabular}{lccc}
\hline
 &
  \multicolumn{1}{c}{\textbf{Exp 1: Sym. MIM}} &
  \multicolumn{1}{c}{\textbf{Exp 2: Asym. MIM}} &
  \multicolumn{1}{c}{\textbf{Exp 3: ES}} \\ \hline

  \multicolumn{1}{l}{\textbf{Metric}} &
  \multicolumn{1}{c}{time to $N_{\text{Bell}}=1000$} &
  \multicolumn{1}{c}{time to $N_{\text{Bell}}=1000$} &
  \multicolumn{1}{c}{fidelity $F$} \\ 

  \multicolumn{1}{l}{\textbf{Node distance} $L$} &
  \multicolumn{1}{c}{20~km} &
  \multicolumn{1}{c}{20~km} &
  \multicolumn{1}{c}{40~km} \\

  \multicolumn{1}{l}{\textbf{BSA location} $d$} &
  \multicolumn{1}{c}{10~km} &
  \multicolumn{1}{c}{$\{10,11,12,\ldots,20\}$~km} &
  \multicolumn{1}{c}{20~km} \\

  \multicolumn{1}{l}{\textbf{Memories/node} $N_{\text{mem}}$} &
  \multicolumn{1}{c}{$\{1,2,4,8,16\}$} &
  \multicolumn{1}{c}{1} &
  \multicolumn{1}{c}{1} \\

  \multicolumn{1}{l}{\textbf{CNOT err. prob.} $p_g$} &
  \multicolumn{1}{c}{-} &
  \multicolumn{1}{c}{-} &
  \multicolumn{1}{c}{$0\leq p_{\text{g}}\leq1$} \\

  \multicolumn{1}{l}{\textbf{Meas. err. prob.} $p_m$} &
  \multicolumn{1}{c}{-} &
  \multicolumn{1}{c}{-} &
  \multicolumn{1}{c}{$0\leq p_{\text{m}}\leq1$} \\

  \multicolumn{1}{l}{\textbf{Mem. coherence time} $\tau$} &
  \multicolumn{1}{c}{-} &
  \multicolumn{1}{c}{-} &
  \multicolumn{1}{c}{$\{$18, 55, $\infty\}$~ms} \\ \hline
\end{tabular}%
}
\vspace{-0.15in}
\end{table}

A natural approach to simulator verification is to compare its predictions with analytical results~\cite{zang2024analytical}, but due to the complexity of quantum networks, it can be done only for very simple cases.
Furthermore, the design and implementation phase of simulator development may potentially introduce subtle assumptions about the theoretical model itself.
To go beyond such limitations and increase confidence in simulation results, we cross-validate QuISP and SeQUeNCe through a set of experiments of increasing complexity.
We concentrate on two main metrics.
The first is the total time each simulator takes to satisfy a particular connection request.
This is meant to cross-validate the probabilistic nature of entanglement generation as well as the implementation and timing of classical messaging.
The second is the fidelity of the end-to-end entanglement.
This will aid in cross-validating the error models in both simulators.

We propose three experiments targeting these two metrics under different scenarios.
The first two experiments focus on link-level entanglement (see Fig.~\ref{fig:preliminaries}(a)) generation by measuring the time required to generate $N_{\text{Bell}}$ Bell pairs.
The third experiment investigates entanglement swapping (see Fig.~\ref{fig:preliminaries}(b)) and the fidelity of the resulting end-to-end Bell pair.

\textit{Experiment 1: Symmetric MIM Link---} BSA is placed in the middle of the MIM link.
Total length of the link is $L=20$~km, meaning the BSA is placed $d=10$~km away from both nodes.
We vary the number of quantum memories, $N_{\text{mem}}\in\{1,2,4,8,16\}$, inside the two quantum nodes, and record the total time required to create $N_{\text{Bell}}=1000$ base Bell pairs.

\textit{Experiment 2: Asymmetric MIM link---} The BSA location is shifted gradually from the middle of the link towards one of the nodes in increments of 1~km until the link becomes a Memory-Memory (MM) link.
The number of quantum memories is fixed to be $N_{\text{mem}}=1$, and we measure the total time required to generate $N_{\text{Bell}}=1000$ base Bell pairs.

\textit{Experiment 3: Entanglement swapping---}
The two end nodes are separated by a distance of 40~km with a repeater node placed in the middle.
The repeater performs ES on the two elementary Bell pairs shared between each end-node and the repeater.
We measure the fidelity of the Bell pairs shared between the two end-nodes after ES.
Sources of noise include imperfect two-qubit CNOT gates, measurement errors, and decoherence of the quantum memories.

Summary of the experiment configurations can be found in Table~\ref{tab:experiments}. The fiber attenuation rate is 0.2~dB/km, the repetition rate is 1~GHz, and the speed of light in fiber is $c=2\times 10^5$~km/s.

\section{Communication Models and Simulated Errors} \label{Models}

We now discuss the different implementations of communication timing and error models in the two simulators.

\emph{Communication model for QuISP---}
\begin{figure}
    \centering
    \includegraphics[width=\linewidth]{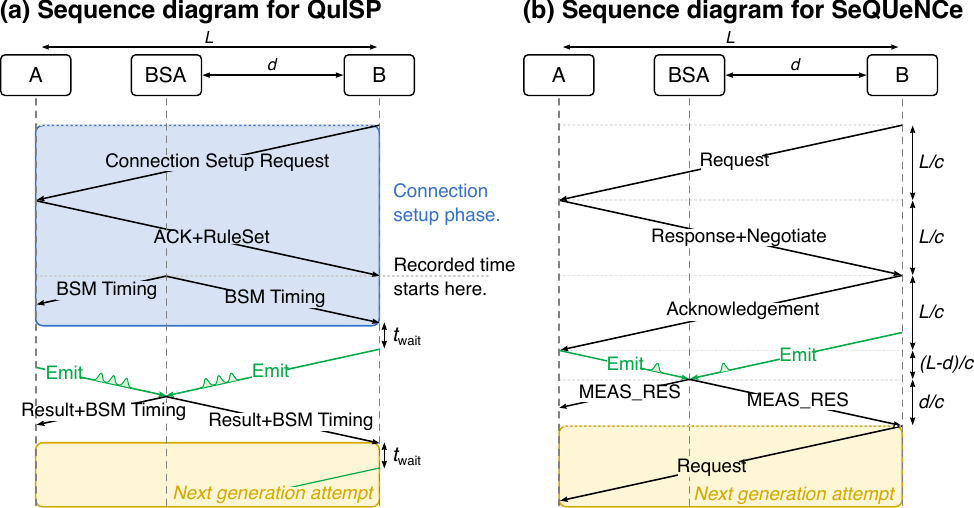}
    \caption{(a) Sequence diagram for link-level entanglement generation in QuISP. The photon train is represented by a single arrow. (b) Timing analysis for the entanglement generation protocol in SeQUeNCe. Classical messages are represented by black arrows and quantum messages by green arrows.}
    \vspace{-0.2in}
\label{fig:sequance_diagram}
\end{figure}
The entanglement generation begins with a connection setup phase, as depicted in Fig.~\ref{fig:sequance_diagram}(a).
Node B sends a connection request specifying $N_{\text{Bell}}$.
Node A replies with an acknowledgment and a generated RuleSet.
At this point ($t=0$), QuISP starts recording the time to generate $N_{\text{Bell}}$ Bell pairs.
Also at $t=0$, the BSA sends a classical message instructing the quantum memories at nodes A and B when to emit photons for synchronization, concluding the connection setup phase.

After the message from the BSA is received by the more distant node (in our case node B), the first Round starts.
The nodes prepare their memories in the wait period $t_{\text{wait}}$, which will emit a train of $N_{\text{mem}}$ photons towards the BSA.
After all measurements, the BSA replies with the measurement results and new timings for the next photonic train.
This concludes one Round and the process is repeated until node B receives the notification from the BSA that the final Bell pair was generated. The generated Bell pairs are immediately consumed to free up memories for further generation attempts.

We denote the expected time to generate $N_{\text{Bell}}$ pairs by $T_{\text{exp}}(N_{\text{Bell}})$.
We assume that the time for the BSA to perform its measurements and classical processing is negligible compared to signal propagation and other operations.
The time for both nodes to receive a message from the BSA is $T_0 = \max(L_A, L_B) / c=d/c$ given our assumption that $d\geq L/2$.
The initial two-way handshake of the connection setup phase takes $2L/c$, while the BSM Timing message takes $T_0$, leading to the total time for the connection setup phase $T_{\text{setup}} = 2L/c + d/c$.
The photons in a train are separated by a time interval $t_{\text{sep}}$.
The waiting time also varies with the quantum technology used for the memories leading it to be also dependent on $t_{\text{sep}}$.
In QuISP, it is set to $t_{\text{wait}}=10t_{\text{sep}}$.
We can now express the duration of one round as $T_{\text{round}} = 2T_0 + 10t_{\text{sep}} + (N_{\text{mem}}-1)t_{\text{sep}}$.
Then the expected number of rounds is $ k = \left\lceil\frac{N_{\text{Bell}}}{N_{\text{mem}}p_{\text{succ}}}\right\rceil$,
where $p_{\text{succ}}$ is the combined probability of both photons arriving at the BSA and undergoing a successful BSM.
This leads to the expected time to generate $N_{\text{Bell}}$ Bell pairs $T_{\text{exp}}(N_{\text{Bell}}) = T_{\text{setup}} + k T_{\text{round}}$.

\emph{Communication model for SeQUeNCe---}
SeQUeNCe also continuously generates entanglement, assuming that an application consumes entanglement immediately upon generation. Nodes A and B negotiate the time and availability of memories for generating entanglement, as shown in Fig.~\ref{fig:sequance_diagram}(b). The protocol is initiated ($t=0$) when a node designated as the primary (e.g., node B) sends a resource manager request to a secondary node (e.g., node A) to generate entanglement. The protocol ends when the BSM has communicated the end of the protocol to both the communicating parties. 
SeQUeNCe restarts the generation protocol for every new Bell pair.

Nodes A and B perform a three-way handshake-like protocol to establish the generation of protocol parameters, taking $3L/c$ seconds.
When multiple memories are available, the protocol is run for each memory pair simultaneously. The photon emissions are time multiplexed into bins which distinguish the transmitted photons according to the pairs of memories communicating with each other. We ignore the time delay introduced due to the time multiplexing when multiple memories operate simultaneously. 
\rdv{Is that the same as or different from QuISP?}
The time required for one round of entanglement generation is $T_{\text{round}} = T_0+d/c+(L-d)/c=4L/c$.
Hence, to generate $N_{\text{Bell}}$ Bell pairs, SeQUeNCe is expected to take $T_{\text{exp}}(N_{\text{Bell}}) = k(4L/c)$,
where $k$ is again the expected number of rounds.
Due to the fact \sequence performs the connection setup phase for every Bell pair, we expect the total time to be slower than \quisp's.

\emph{Error models for QuISP---}
QuISP treats time-dependent errors by discrete exponentiation of a time-invariant state transition matrix.
Time $t$ is approximated by $n=\lceil t/\delta t\rceil$ slices of duration $\delta t$.
QuISP associates each qubit with an \emph{error vector probability} $\vec{\pi}(t) = (\pi_I, \pi_X, \pi_Y, \pi_Z, \pi_R, \pi_E, \pi_L)$,
where $\pi_j$ represents the probability of the qubit being in the state affected by error $j$ at time $t$, and $\sum_j \pi_j = 1$.
Errors $\{X,Y,Z\}$ represent Pauli errors, $\{R,E\}$ are relaxation and excitation errors, and $L$ is photon loss.
The error probability vector evolution is described by a \emph{transition matrix} $Q$, s.t. the error vector probability at time $t$ is given by $\vec{\pi}(t) = \vec{\pi}(t - \delta t) Q = \vec{\pi}(0) Q^n$.
QuISP allows users to specify the transition matrix $Q$.
Before any quantum operation, \quisp samples the error probability vector turning the qubits into a definite pure state known to the simulator but not to the network protocols and applications being modeled.

For single-qubit measurements, \quisp performs a perfect measurement, and with error probability $p_m$ flips the reported measurement outcome.
For two-qubit gates, \quisp first applies the noiseless gate followed by applying a gate error with probability $p_g$. 
If the gate is determined to be noisy, a two-qubit Pauli operator is sampled from the set of 15 possible errors $(\{I, X, Y, Z\} \otimes \{I, X, Y, Z\} - \{I\otimes I\})$ with user specified weights.

Assuming that link-level Bell pairs are noiseless, the fidelity of the end-to-end Bell pair is
\begin{align}
    F_\mathrm{swap} & = (1 - p_{\text{g}})  (1 - p_{\text{m}})^2 + \frac{3}{15} p_{\text{g}}  (1 - p_{\text{m}})^2 \nonumber\\
        &+ \frac{8}{15} p_{\text{g}}  (1 - p_{\text{m}}) p_{\text{m}} + \frac{4}{15} p_{\text{g}}  p^2_{\text{m}}.
\label{eq:quisp-swap-fidelity-perfect-memory}
\end{align}
Taking memory decoherence into account, we can modify Eq.~(\ref{eq:quisp-swap-fidelity-perfect-memory}) as
\begin{align}
    F_\mathrm{swap, decoherence} = F_\mathrm{swap} \cdot  \left(Q^{2t_1 + 2t_2 + 2T}\right)_{00},
    \label{eq:quisp-swap-fidelity-with-decoherence}
\end{align}
with $t_1$ ($t_2$) being the time since photon emission until link-level Bell pair is created for A (B), and $T$ being the time it takes for the swap message to arrive at A and B from the repeater. 
For depolarizing errors, we have a closed form expression for any integer power of the transition matrix $\left(Q^t_\mathrm{depol}\right)_{00} = \left[1+3^{1-t}(3-4 p)^t\right]/4$,
where $p$ is chosen such that $\left(Q^{t_\mathrm{coherence}}_\mathrm{depol}\right)_{00} = 1/e$.

\begin{figure}
    \centering
    \includegraphics[width=\columnwidth]{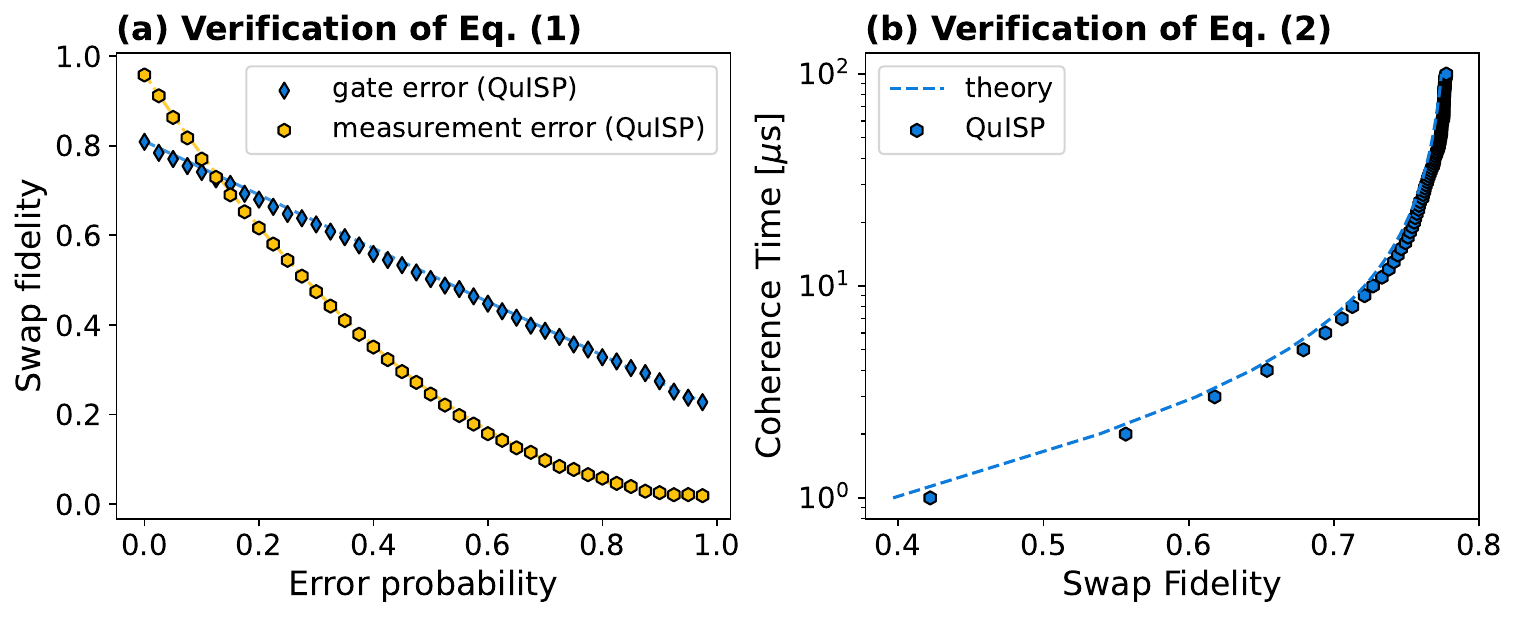}
    \caption{(a) Verification of end-to-end fidelity Eq.~(\ref{eq:quisp-swap-fidelity-perfect-memory}) for QuISP. Blue diamonds for varying $p_{\text{g}}$ and $p_{\text{m}}=0.1$ and yellow hexagons for varying $p_{\text{m}}$ with $p_{\text{g}}=0.05$. Dashed lines represent our theoretical predictions. (b) Verification of Eq.~(\ref{eq:quisp-swap-fidelity-with-decoherence}) for $p_{\text{g}}=0.05$ and $p_{\text{m}}=0.1$, varying degree of depolarizing noise.}
    \label{fig:quisp-error-against-theoretical}
\end{figure}

Figure~\ref{fig:quisp-error-against-theoretical}(a) shows the end-to-end fidelity as a function of varying either the gate error or the measurement error with the other error rate being fixed.
Our predictions using Eq.~(\ref{eq:quisp-swap-fidelity-perfect-memory}) display excellent agreement with \quisp.
In Fig.~\ref{fig:quisp-error-against-theoretical}(b), we include the effect of depolarizing noise while fixing both gate and measurement errors to a finite value and again observe agreement between Eq.~(\ref{eq:quisp-swap-fidelity-with-decoherence}) and the ouput of the simulator.

\emph{Error models for SeQUeNCe.---}
\begin{figure}
    \centering
    \includegraphics[width=\columnwidth]{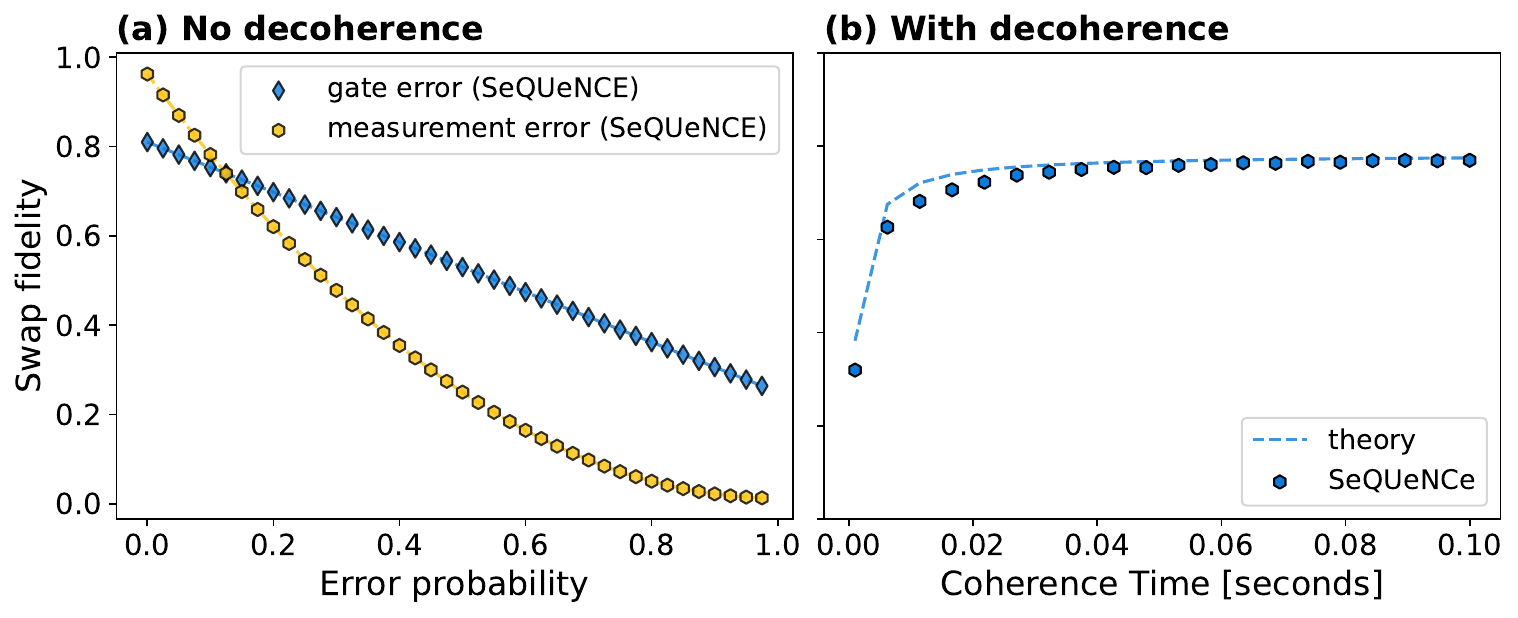}
    \caption{Verification of \sequence error models. (a) Without decoherence. Blue diamonds for $p_{\text{m}}=0.1$ and yellow hexagons for $p_{\text{g}}=0.05$. (b) Including decoherence for $p_{\text{g}}=0.05$ and $p_{\text{m}}=0.1$.}
    \vspace{-0.2in}
    \label{fig:sequence_fidelity_verification}
\end{figure}
We now specify the error models in SeQUeNCe. Quantum states stored in quantum memories decohere over time, and the quantum gates and measurements in entanglement swapping are imperfect.

For quantum memory decoherence, SeQUeNCe implements the analytical model of continuous-time Pauli channel. We focus on the depolarizing channel and entanglement in the form of a \emph{Werner state}. The effect on a single qubit is to preserve the quantum state with probability $p$ and to bring the state into a maximally mixed state with probability $(1-p)$. We can model the time-dependence of the depolarizing channel with $p=\exp(-t/\tau)$, where $\tau$ is the memory coherence time. 
If a Werner state of fidelity $F_{\text{in}}$ is stored in two quantum memories with identical coherence times $\tau$ for time $t$, the final state is a Werner state with degraded fidelity~\cite{zang2023entanglement} $F(t) = F_{\text{in}} e^{-2t/\tau} + (1-e^{-2t/\tau})/4$.
For two qubit-gates, with probability $(1-p_{\text{g}})$ (gate fidelity) the expected gate operation is applied noiselessly and with probability $p_{\text{g}}$ the two involved qubits are completely depolarized.
For single-qubit measurements in the computational basis,  $(1-p_{\text{m}})$ is the probability to get the correct outcome, and with probability $p_{\text{m}}$ the measurement outcome is flipped.

We consider ES with noisy CNOT gate and imperfect single-qubit measurements, with two Werner states as input. The output fidelity is~\cite{zang2023entanglement}
\begin{align}
    F_\mathrm{swap,W} & = \frac{p_{\text{g}}}{4} + (1-p_{\text{g}})\left[(1-p_{\text{m}})^2\left(F_1F_2 + 3e_1e_2\right)\right.\nonumber \\
    & \left. +p_{\text{m}}(p_{\text{m}}-2)\left(F_1e_2 + e_1F_2 + 2e_1e_2\right)\right],
    \label{eq:sequence-fidelity-verification}
\end{align}
where $F_1,F_2$ are the fidelities of the input Werner states, and we have defined $e_i=(1-F_i)/3$ to simplify the notation.

We verify our model for imperfect operations in Fig.~\ref{fig:sequence_fidelity_verification}(a).
We observe excellent agreement between Eq.~(\ref{eq:sequence-fidelity-verification}) and the output of \sequence when the two-qubit gates or the measurements are noisy.
Effect of decoherence in the form of depolarizing noise is depicted in Fig.~\ref{fig:sequence_fidelity_verification}(b).
Again, we observe good agreement between our model and simulator.

\section{Results}
\label{Results}

\begin{figure}
    \centering
    \includegraphics[width=\columnwidth]{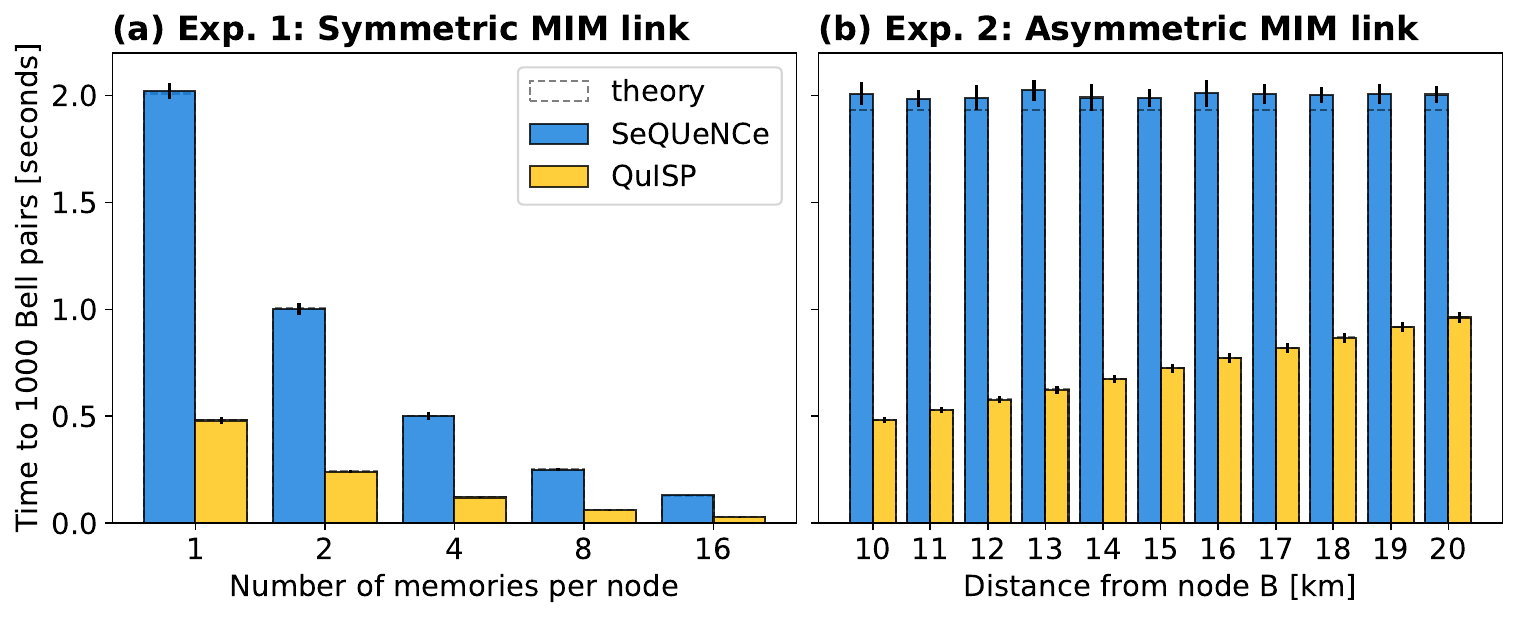}
    \caption{Time to generate 1000 Bell pairs for symmetric MIM link in (a), and asymmetric MIM link in (b). QuISP simulation used $t_{\text{sep}}=1$~ns.}
    \vspace{-0.2in}
    \label{fig:MIM_time}
\end{figure}

\emph{Symmetric MIM link.---} Results of Experiment 1 are shown in Fig.~\ref{fig:MIM_time}(a), where the total time to generate 1000 Bell pairs is plotted against the number of quantum memories available at each node.
Dashed line represents theoretic predictions for QuISP and SeQUeNCe.
Both simulators agree excellently with their respective theoretical models, and also with each other on the general qualitative behavior that utilizing multiple memories leads to faster distribution of the requested number of Bell pairs.
However, the simulators disagree on the actual time it takes to do so, with SeQUeNCe being slower than QuISP.
This is explained by the two different approaches that the simulators take when setting up the connection between A and B (refer back to Section~\ref{Models}).
SeQUeNCe performs a three-way handshake for every single Bell pair as explained in Fig.~\ref{fig:sequance_diagram}(b) while QuISP performs a two-way handshake only once.
One would thus expect the ratio of the two times to be roughly constant for varying number of memories.
This is indeed the case, with the ratio varying only slightly between 4.16 and 4.33. 

\emph{Asymmetric MIM Link.---} The case of the asymmetric MIM link demonstrated in Fig.~\ref{fig:MIM_time}(b) shows qualitative difference between the simulators.
The time it takes to setup a connection in SeQUeNCe is independent of the actual placement of the BSA.
This is evident also from the theoretical expected time $T_{\text{exp}}(N_{\text{Bell}})$. 
QuISP's connection setup protocol on the other hand performs best when the BSA is located in the middle, and its performance deteriorates as the distance between the BSA and node B increases.

\begin{figure}
    \centering
    \includegraphics[width=\columnwidth]{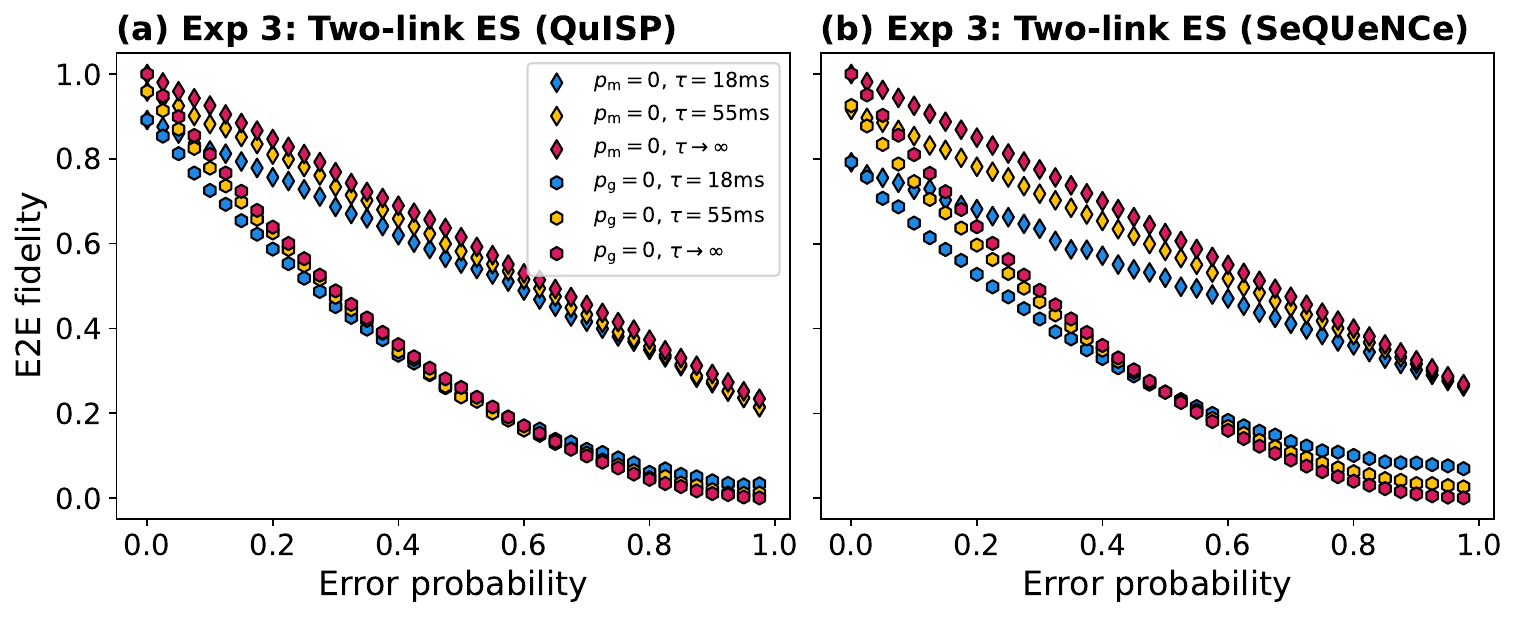}
    \caption{ES under noisy operations and memory decoherence. Diamonds represent perfect measurements, hexagons ideal gates.}
    \vspace{-0.2in}
    \label{fig:exp3_es}
\end{figure}

\emph{Two-link entanglement swapping.---} Figure~\ref{fig:exp3_es} shows the results for the case of two-link entanglement swapping under the influence of noisy gates, measurements, and depolarizing quantum memories.
The diamonds represent end-to-end fidelities for perfect measurements, while the hexagons represent the case of perfect two-qubit gates.
The colors represent varying coherence times with $\tau\rightarrow\infty$ meaning no decoherence.

Both simulators agree that faulty measurement devices are more detrimental to the end-to-end fidelity than gate errors with comparable rates.
The simulators disagree on the quantitative behavior of end-to-end fidelity, particularly for quantum memories with poor coherence times, with \quisp predicting higher values for low error probability range while predicting lower values for higher error probability range.
This can be traced back to the discussed differences in the simulators' communication models and their slight difference in the two-qubit gate error description. 
\quisp is generally faster when it comes to generation of link-level entanglement, as seen in Fig.~\ref{fig:MIM_time}(a).
So the link-level Bell pairs spend less time in the memories before undergoing entanglement swapping. Higher base fidelities subsequently lead to higher end-to-end quality of the distributed Bell pairs.

\section{Discussion}
\label{Discussion}

We have performed a cross-validation of \quisp and \sequence, two prominent open-source quantum network simulators.
We studied the link-level entanglement generation to evaluate how long each simulator takes to satisfy a connection request.
\quisp relies on a protocol requiring fewer classical messages, leading to faster rate of link-level entanglement compared to \sequence.
Interestingly, we discovered that the slower connection setup protocol for \sequence is insensitive to the placement of the BSA node, a property that might be of interest when planning the deployment of first-generation quantum networks.
We also focused on how the simulators handle noisy operations and decoherence of quantum memories.
The simulators agreed on the general qualitative behavior but differed on the specific values of end-to-end fidelity.
This can be understood by the different communication models. 

The tests presented here have strengthened our confidence in both simulators and deepened our understanding of the impact on performance given differences in protocols. The tests described in this paper are deliberately chosen to also be analytically tractable, allowing us to write down closed-form equations, which match the simulation outcomes.  

The obvious next step is to move beyond analytically tractable cases and compare the results of more complex systems. 
The interplay between differences in both communication models and error models will be explored in more detail as future work.
Moreover, first-generation quantum networks use \emph{purification} or \emph{distillation}. The details in the operation will result in a substantially extended analysis.

Despite the existence of suites of software quality assurance (SQA) tests for both simulators, this exercise has not only given us a deeper understanding of the network design choices but also resulted in numerous outright bug fixes to both simulators. We recommend this direct comparison approach for both quantum and classical network simulators where possible.

\bibliographystyle{IEEEtran.bst}
\bibliography{bibfile}

\end{document}

%% file: custom-command.tex
\usepackage[svgnames, table]{xcolor}
\usepackage{tabularx, makecell, linegoal}

\usepackage{ifthen}

\newboolean{ShowComments}
\setboolean{ShowComments}{false}  
\ifthenelse{\boolean{ShowComments}}%
	{
		\newcommand{\ColorComment}[3]{%
				{\colorbox{#1}{\color{White}   \textsf{\textbf{#2}}} \textcolor{#1}{#3}}}

	}%
	{
		\newcommand{\ColorComment}[3]{}

	}%

\definecolor{rdvcolor}{rgb}{0,0.5,0}\newcommand{\rdv}[1]{\ColorComment{rdvcolor}{rdv}{#1}}
\definecolor{michalcolor}{RGB}{255,127,80}
\definecolor{naphancolor}{RGB}{165,6,245}
\definecolor{joaquincolor}{rgb}{0,0.5,0.8}
\definecolor{simoncolor}{rgb}{0,0.5,0}
\definecolor{allencolor}{RGB}{254,0,0}
\definecolor{alexcolor}{RGB}{75,44,249}

\newcommand{\quisp}{QuISP\xspace}
\newcommand{\sequence}{SeQUeNCe\xspace}